\documentstyle[prl,multicol,aps]{revtex}

\begin{document}
\draft

\title{Microscopic analysis of a correlation between dipole
transitions $1^-_1 \to 0^+_{g.s.}$ and $3^-_1 \to 2^+_{1}$
in spherical nuclei}
\author{V.Yu.~Ponomarev\thanks{NATO fellow. Permanent address: Bogoliubov 
Laboratory of Theoretical Physics, Joint Institute for
Nuclear Research, 141980, Moscow region, Dubna, Russia, 
E-mail: vlad@thsun1.jinr.ru
}}
\address{University of Gent, Department of Subatomic and Radiation
Physics, Proeftuinstraat 86, 9000 Gent, Belgium}
\date{\today }
\maketitle

\begin{abstract}
Correlation between B($E1, 1^-_1 \to 0^+_{g.s.}$) and
B($E1, 3^-_1 \to 2^+_{1}$) values are considered within microscopic
QRPA approach. 
General arguments for a dependence of a ratio between these values
on a collectivity of the $2^+_1$ and $3^-_1$ phonons and ground state
correlations are provided.
\end{abstract}

\begin{multicols}{2}
\narrowtext

Recently, impressive correlation between B($E1, 1^-_1 \to 0^+_{g.s.}$) 
and B($E1, 3^-_1 \to 2^+_{1}$) values have been reported by
N. Pietralla \cite{Pie99} from an experimental systematic.
Although the measured B($E1$) values themselves vary within two orders of
magnitude for different medium-heavy spherical nuclei, the ratio
\begin{equation}
R =  \frac{\mbox{B}(E1, 1^-_1 \to 0^+_{g.s.})}
{\mbox{B}(E1, 3^-_1 \to 2^+_{1})} 
\label{R}
\end{equation}
keeps practically constant and is close to one. 
In the same paper it is shown that $R=7/3$ if a simple bosonic phonon 
model is applied to to describe the states involved. 
In this Brief report we present a microscopic analysis
of transition matrix elements under consideration and demonstrate that
$0<R<7/3$ in any theoretical model based on quasiparticle random phase
approximation (QRPA) approach if internal fermion structure of phonons is 
accounted for.

Let us introduce a phonon operator ($Q^{+}_{\lambda \mu }$) 
with a multipolarity $\lambda$ and projection $\mu$ 
to describe excited states in
nuclei as a superposition of different two-quasiparticle
configurations:
\begin{eqnarray}
Q^{+}_{\lambda \mu }\,&=&\,\frac{1}{2}
\sum_{\tau}^{n,p} \sum_{jj'} \left \{ 
C_{jm j'm'}^{\lambda \mu} 
X_{jj'}^{\lambda } 
\alpha^{+}_{jm} \alpha^{+}_{j'm'}
 \right .\nonumber
 \\
&-&\left .(-1)^{\lambda - \mu} 
C_{j'm' jm}^{\lambda -\mu} Y_{jj'}^{\lambda }
\alpha_{j'm'} \alpha_{jm}
\right \}~.
\label{ph}
\end{eqnarray}
The quantity $jm \equiv |nljm>$ denotes a single-particle level of an
average field and $C$ is a Clebsh-Gordon coefficient.
Quasiparticle operators, $\alpha_{jm}^{+}$, are obtained from
a linear Bogoliubov transformation from
the particle creation $a_{jm}^{+}$ and annihilation $a_{jm}$ operators:
\begin{eqnarray}
a_{jm}^+ &=& u_j\alpha _{jm}^+ +(-1)^{j-m}v_j\alpha _{j-m}   \nonumber \\
a_{jm}   &=& u_j\alpha _{jm}   +(-1)^{j-m}v_j\alpha _{j-m}^+ \nonumber
\end{eqnarray}
where $u_j^2$ and $v_j^2$ are occupation numbers.

The properties of phonons of Eq.~(\ref{ph}), 
i.e. their excitation energies
($E_{\lambda}$) and the values of forward ($X^{\lambda}_{j j'}$) 
and backward ($Y^{\lambda}_{j j'}$) amplitudes are obtained by
solving the QRPA equations. We are interested here only in the first
collective QRPA solution, thus, index $i$ which is often used to
distinguish phonons with a different excitation energy will be dropped.

It is well-known that in spherical nuclei the lowest $2^+$ and $3^-$
excited states have a practically pure one-phonon nature while
the lowest $1^-$ one is a two-phonon state which we describe by a
wave function:
\[
|1^-_1 \mu_1> = \sum_{\mu_2 \mu_3} C_{2 \mu_2 3 \mu_3}^{1 \mu_1}
Q^{+}_{2^+ \mu_2} Q^{+}_{3^- \mu_3} |>_{ph}
\] 
where $|>_{ph}$ is a wave function of a ground state of
an even-even nuclei, a phonon vacuum.

In terms of quasiparticles and phonons, a one-body operator of an 
electromagnetic $E\lambda$ transition has the form:
\begin{eqnarray}
&&{\cal M}(E \lambda \mu ) =
\sum_{\tau}^{n,p} e^{(\lambda)}_{\tau}
\sum_{j j'} \frac{<j||E \lambda||j'>}
{\sqrt{2 \lambda +1}}  \nonumber \\ 
&&\times \left \{
\frac{(u_j v_{j'} + v_j u_{j'})}{2} 
(X^{\lambda }_{j j'} + Y^{\lambda }_{j j'})
(Q^{+}_{\lambda \mu }+(-)^{\lambda -\mu }Q_{\lambda -\mu })
\right .     \nonumber
\\
&&+\left . (u_j u_{j'} - v_j v_{j'}) 
\sum_{mm'}C_{jmj'm'}^{\lambda \mu }(-)^{j'+m'}
\alpha_{j m}^{+}\alpha _{j'-m'}
\right \}
\label{tr2}
\end{eqnarray}
where $<j||\mbox{E} \lambda||j'> \equiv 
<j||i^{\lambda} Y_{{\lambda}} r^{\lambda}||j'>$  
is a single particle transition matrix element 
and $e^{(\lambda)}_{\tau}$ are effective charges for neutrons and
protons.
The first term of Eq.~(\ref{tr2}) corresponds to a one-phonon exchange
and it does not contribute for transitions between the one-phonon
$3^-_1$ and $2^+_1$ excited states and for a decay of the two-phonon
$1^-_1$ state into the ground state.

Applying exact commutation relations between pho\-non and quasiparticle 
operators:
\begin{eqnarray}
\nonumber 
[\alpha_{jm}, Q^+_{\lambda \mu } ]_{\_} &=& \sum_{j'm'}
X^{\lambda}_{jj'} C^{\lambda \mu}_{jm j'm'} \alpha^+_{j'm'}~,
\\
\label{a-q-c}
[\alpha^+_{jm}, Q^+_{\lambda \mu } ]_{\_} &=&
(-1)^{\lambda - \mu} \sum_{j'm'}
Y^{\lambda}_{jj'} C^{\lambda -\mu}_{jm j'm'} \alpha_{j'm'}
\nonumber
\end{eqnarray}
we obtain
\begin{eqnarray}
&&\mbox{B}(E \lambda_1 ; 
[\lambda_2 \times \lambda_3 ]_{\lambda_1}
 \rightarrow 0^{+}_{g.s.}) =
\frac{(2\lambda_2+1)(2\lambda_3+1)}{(2\lambda_1+1)}   
\nonumber
 \\
\times&& \left |
\sum_{\tau}^{n,p} e^{(\lambda_1)}_{\tau}
\sum_{j_{1} j_{2} j_{3}}  (u_{j_1} u_{j_2}- v_{j_1} v_{j_2})
<j_1 ||\mbox{E}\lambda_1 || j_2>
 \right .
\nonumber \\
\times&&
\left .
\left\{
\begin{array}{ccc}
\lambda_3 & \lambda_2 & \lambda_1 \\
j_{1}  & j_{2}  & j_{3} \end{array} \right\}
\left( X^{\lambda_3}_{j_{2} j_{3}}  \:
Y^{\lambda_2}_{j_{3} j_{1}} +
Y^{\lambda_3}_{j_{2} j_{3}} \:
X^{\lambda_2}_{j_{3} j_{1}}   
\right) \right | ^{2} 
\label{qq-gs}
\end{eqnarray}
for the $E1$-decay $1^-_1 \to 0^+_{g.s.}$ and 
\begin{eqnarray}
&&\mbox{B}(E\lambda_1, \lambda_3 \rightarrow
\lambda_2 ) = (2\lambda_2+1)
\nonumber
 \\
\times&&
\left | \sum_{\tau}^{n,p} e^{(\lambda_1)}_{\tau}
\sum_{j_{1} j_{2} j_{3}}  (u_{j_1} u_{j_2}- v_{j_1} v_{j_2})
<j_1 ||\mbox{E}\lambda_1 || j_2>
 \right .
\nonumber \\
\times&&
\left .
\left\{
\begin{array}{ccc}
\lambda_3 & \lambda_2 & \lambda_1 \\
j_{1}  & j_{2}  & j_{3} \end{array} \right\}
\left( X^{\lambda_3}_{j_{2} j_{3}}  \:
X^{\lambda_2}_{j_{3} j_{1}}   +
Y^{\lambda_3}_{j_{2} j_{3}}  \:
Y^{\lambda_2}_{j_{3} j_{1}}
\right) \right | ^{2} 
\label{q-q}
\end{eqnarray}
for the $E1$-decay $3^-_1 \to 2^+_{1}$, where $\lambda_1 = 1$, 
$\lambda_2^{\pi} = 2^+$ and $\lambda_3^{\pi} = 3^-$ .

Assuming $X^{\lambda}_{j j'} \equiv Y^{\lambda}_{j j'}$ and
equal effective charges ($e^{(\lambda_1)}_{\tau}$) for both $E1$ 
transitions under consideration, Eqs. (\ref{qq-gs},\ref{q-q}) 
yield the value $R=7/3$, the same as in 
a simple bosonic phonon model \cite{Pie99}.
In fact, the amplitudes $X^{\lambda}_{j j'}$ are always larger
than $Y^{\lambda}_{j j'}$ amplitudes. For example, in models
with a separable form of a residual interaction they have the
following analytical expressions: 
\begin{equation}
{X \choose Y}^{\lambda}_{j j'} (\tau) =
\frac{1} {\sqrt{ \cal{Y} _{\tau} ^{\lambda } }}
\cdot
\frac{f^{\lambda}_{j j'}(\tau) (u_j v_{j'}+ u_{j'} v_j)}
{\varepsilon_{jj'} \mp E_{\lambda }}
\label{pp}
\end{equation} 
where $\varepsilon_{jj'}$ is an energy of a two-quasiparticle 
configuration ($\alpha^+_{j} \alpha^+_{j'}$), 
$f_{j j'}^{\lambda}$ is a reduced single-particle matrix element of
residual forces,
and the value $\cal{Y} _{\tau} ^{\lambda}$ is determined from
a normalization condition for phonon operators:
\[
\langle | Q_{\lambda \mu} Q_{\lambda \mu}^+ | \rangle_{ph} =
 \sum_{\tau}^{n,p} \sum_{jj'} \left \{ ( X^{\lambda}_{jj'})^{2} -
(Y^{\lambda}_{jj'} \right )^{2} \}  = 1~.
\]
As we notice from Eq.~(\ref{pp}), $X^{\lambda}_{j j'}$ and
$Y^{\lambda}_{j j'}$ amplitudes always have the same sign for the first
collective phonon because $E_{\lambda} < \varepsilon_{jj'}$. 
An approximation 
$X^{\lambda}_{j j'} \approx Y^{\lambda}_{j j'}$ 
is valid only when the phonon energy is very small as compared to
two-quasiparticle energies, i.e. $E_{\lambda} << \varepsilon_{jj'}$,
and corresponds to extremely collective vibrations.
Thus, the value 7/3 should be considered as an upper unreachable 
limit for the quantity $R$ of Eq.~(\ref{R}).

Since it is always true that 
$X^{\lambda}_{j j'} > Y^{\lambda}_{j j'}$,
all elements of sum in Eq.~(\ref{qq-gs}) are systematically smaller
than the corresponding ones in Eq.~(\ref{q-q}) reducing the value of
$R$ from the upper 7/3 limit.
In spherical nuclei the excitation energies of the lowest vibrational
$2^+$ and $3^-$ states approximately equal to $2/3$  
of an energy of the lowest two-quasiparticle configuration 
($\varepsilon_{(jj')_{l}}$).
Keeping only the main term in sums of Eqs.~(\ref{qq-gs},\ref{q-q})
we obtain the value $R \approx 0.9$.
In fact, the quantity $R$ should be somewhat larger because
the ratios $X^{2(3)}_{j j'}/Y^{2(3)}_{j j'}$ for
omitted terms are smaller than for the main, 
$X^{2(3)}_{(jj')_l}/Y^{2(3)}_{(jj')_l}$, ones.

While $R=7/3$ should be taken as an upper extreme limit as discussed
above, another extreme limit is $R=0$. It corresponds to a situation 
when the ground state of the nucleus is considered as a non-correlated 
vacuum in respect to phonon excitations. This is the case when nuclear
excitations are treated within the Tamm-Dankoff approximation (TDA)
approach. Within the TDA the
nucleus ground state is supposed to be a quasiparticle vacuum and the
coefficients $Y^{\lambda}_{j j'} \equiv 0$. It means that a direct
transition $[2^+_1 \times 3^-_1]_{1^-} \to 0^+_{g.s.}$ is completely
forbidden (see, Eq.~(\ref{qq-gs})). It is different in the QRPA
approach because
a direct decay of a two-phonon state into the ground state by means
of one-body operator of electromagnetic transition takes place  
by a simultaneous annihilation of a two-quasiparticle configuration of an
excited state and a virtual excitation of another two-quasiparticle 
configuration in the correlated phonon vacuum.
As for the $3^-_1 \to 2^+_1$ decay between one-phonon states, one
quasiparticle in the $3^-$ phonon simply re-scatters into 
the $2^+$ phonon by
means of an $E1$ operator. It means that the last transition is
allowed in second order perturbation theory in both TDA and QRPA 
approaches. 

The lower limit $R \approx 0$ may be also approached when $E_{\lambda}
\approx \epsilon_{(jj')_{l}}$. This is the case of non-collective
excitations when 
$X_{(jj') \ne (jj')_l}, Y_{(jj') \ne (jj')_l} \approx 0$
and $X_{(jj')_l} >> Y_{(jj')_l}$. It does not take place in atomic nuclei.

Thus, the experimental systematic in Ref.~\cite{Pie99} for the value of
$R \approx 1$ provides us an additional good evidence that in
spherical nuclei the lowest $2^+$ and $3^-$ states are good collective 
vibrators built on top of a correlated vacuum. 

The present paper was partly supported by the Research Council of the
University of Gent and the Russian Fund for Basic Research
(grant No. 96-15-96729).

\end{multicols}

\end{document}